\documentclass{aa}
\usepackage{graphicx}
\usepackage{txfonts}
\begin{document}

    \title{The unusual pre-main-sequence star \mbox{V718 Per} (HMW 15)}
    \subtitle{Photometry and spectroscopy across the eclipse\thanks{Based
      on observations collected at the Crimean Astrophysical Observatory, Ukraine; the W. M. Keck Observatory, Hawaii, USA and the Campo Imperatore Observatory, Italy} }

    \titlerunning{ }

    \author{V. Grinin\inst{1}
    \and H. C. Stempels\inst{2}
    \and G. F. Gahm\inst{3}
    \and S. Sergeev\inst{4}
    \and A. Arkharov\inst{1}
    \and O. Barsunova\inst{1,5}
    \and L. Tambovtseva\inst{1}
    }

    \offprints{V. P. Grinin}

    \institute{
    Pulkovo Astronomical Observatory, 196140,
    Pulkovskoe shosse 65, St.\,Petersburg, Russia\\
    \email{grinin@gao.spb.ru}
    \and
    School of Physics \& Astronomy, University of St. Andrews,
    North Haugh, St Andrews KY16 8RQ, Scotland\\
    \email{Eric.Stempels@st-andrews.ac.uk}
    \and
    Stockholm Observatory, AlbaNova University Centre, SE - 106 91 Stockholm, Sweden\\
    \email{gahm@astro.su.se}
    \and
    Crimean Astrophysical Observatory, Crimea, Nauchny, Ukraine\\
    \email{sergeev@crao.crimea.ua}
    \and
    Sobolev Astronomical Institute, St. Petersburg State University,
    St. Petersburg, Russia
    }

    \date{Received\,/\,Accepted}

\authorrunning{V. Grinin et al.}
\titlerunning{The unusual pre-main-sequence star \mbox{V718 Per}}

\date{Received date; Accepted date}  

\abstract
{The remarkable pre-main-sequence object \mbox{V718 Per} (HMW 15, H187) in the young cluster IC
348 periodically undergoes long-lasting eclipses caused by variable amounts of
circumstellar dust in the line-of-sight to the star. It has been speculated that
the star is a close binary and similar to another unusual eclipsing object, KH
15D.}
{We have submitted \mbox{V718 Per} to a detailed photometric and spectroscopic study to
investigate how regular the recurrent eclipses are, to find out more about
the properties of the stellar object and the occulting circumstellar material,
and to look for signatures of a possible binary component.}
{\mbox{V718 Per} was monitored photometrically from the optical to the
near-infrared (NIR). We also obtained high-resolution optical spectra
with the Keck telescope at minimum as well as at maximum brightness. We derived
the fundamental photospheric parameters of this star by comparing with synthetic
spectra.}
{Our photometric data show that the eclipses are very symmetric and persistent,
and that the extinction law of the foreground occulting dust deviates only
little from what is expected for ``normal'' interstellar material.
The stellar parameters of \mbox{V718 Per}
indicate a primordial abundance of Li and a surface temperature of
$T_{\rm eff} \approx 5200$ K. Remarkably, the in-eclipse spectrum shows a
significant broadening of the photospheric absorption lines, as well as a
slightly lower stellar surface temperature. In addition, weak emission
components appear in the absorption lines of H$\alpha$ and the Ca II IR triplet
lines. We did not detect any signs of atomic or molecular features related to
the occulting body in the in-eclipse spectrum. We also found no evidence of any
radial velocity changes in \mbox{V718 Per} to within about $\pm 80$ m s$^{-1}$,
which for an edge-on system corresponds to a maximum companion mass of
${\sim}6~M_{\rm Jup}$.} 
{Our observations suggest that \mbox{V718 Per} is a single star, and thus very
different from the related binary system KH 15D. We conclude that \mbox{V718 Per} is
surrounded by an edge-on circumstellar disk with an irregular mass distribution
orbiting at a distance of 3.3 AU from the star, presumably at the inner disk
edge. To produce the prolonged eclipses, the occulting feature must extend along
more than half of the inner disk edge. The change in stellar surface temperature
and the emission line activity observed could be related to spot activity. We
ascribe the broadening of photospheric absorption lines during the eclipse to
forward scattering of stellar light in the circumstellar dust feature.}


\maketitle  


\section {Introduction}

\object{V718 Per}, also known as HMW\,15 and H\,187\footnote{Other
designations: TJ 108, LDL 35, LRL 35}, is a late-type pre-main-sequence
star in the nearby, young
open cluster IC 348.  Comprehensive photometric monitoring of \mbox{V718 Per} by Cohen
et al. (\cite{coh03}) revealed a very unusual eclipse in this object. They
detected an extremely long and smooth eclipse ($t_{ecl} \approx$ 3.5 yrs and an
eclipse depth of $\sim 0.7^{\rm m}$), and earlier scattered photometric data
suggested that these eclipses may be recurrent (see Cohen et al.
\cite{coh04}). Late 2004, \mbox{V718 Per} entered a second eclipse, in shape and depth very
similar to the first one (Barsunova et al. \cite{bar05}).
More detailed observations by Nordhagen et al. (\cite{nor06}) show
that \mbox{V718 Per} undergoes recurrent, 3.5-year long eclipses with a
period of $P = 4.7 \pm 0.1$ years (see also Grinin
et al. \cite{griB06}). Thus, given the very long eclipse duration and
its comparitively short period, this system is one of the most exotic eclipsing
systems known. The closest analog to \mbox{V718 Per} is probably the system KH
15D, a weak-line T Tauri binary system that experiences eclipses by its
circumbinary disk
(Kearns \& Herbst \cite{kearns98}; Hamilton et al. \cite{hamilton01};
Winn et al. \cite{win06}).

The extremely long duration of the eclipse, combined with the comparatively
short period rules out periodic eclipses by a compact body, like a stellar or
planetary companion, as the possible cause of variability. Since a 4.7-year
period suggests orbital motion as the underlying cause, most working hypotheses
propose that the eclipses are produced by an extended body, like an irregular or
warped circumstellar or circumbinary disk seen nearly edge-on (Cohen et al.
\cite{coh04};  Nordhagen et al. \cite{nor06}; Grinin et al. \cite{griB06}). A
still unresolved question is whether or not \mbox{V718 Per} is a close binary, similar
to KH 15D, where the orbital displacement of the stars would periodically hide
one or both components behind a shared circumbinary disk. The fact that the
spectral types assigned to the object cover a rather large range, G8 -- K6
(Herbig \cite{her98}; Luhman et al. \cite{luh98}), and may even change with
wavelength, could flag the presence of two stellar spectra. As pointed out by
Nordhagen et al. (\cite{nor06}), a variable radial velocity with an amplitude of
several km s${}^{-1}$ should then be detectable from high-resolution
spectroscopy.

In this paper we present the results of our photometric and spectroscopic
observations of \mbox{V718 Per}. We have obtained multi-waveband photometry, as well as
two high-resolution spectroscopic observations with the Keck telescope. These
observations, together with existing infrared photometry, provide us with a
basis for improving our understanding of the nature of this object and its
unusual eclipses, and we can address the question of binarity of \mbox{V718 Per}.

\begin{figure}
\centering
\includegraphics[angle=00,width=6.5cm]{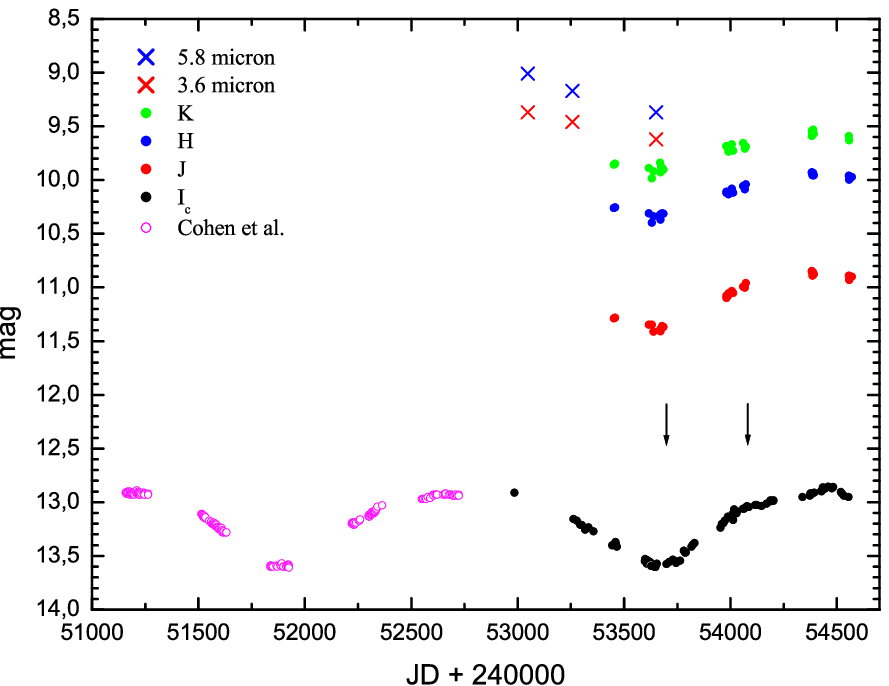}
\includegraphics[angle=00,width=6.5cm]{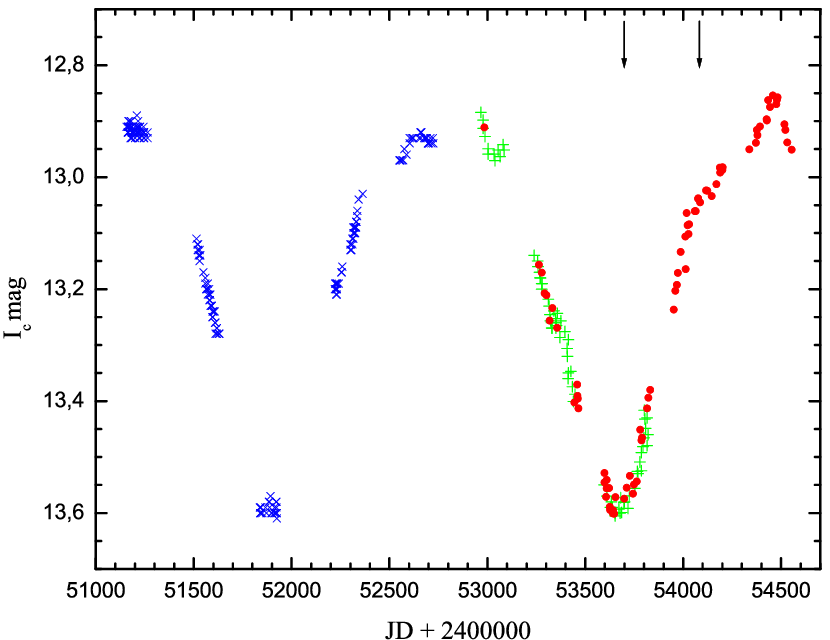}
\includegraphics[angle=00,width=6.5cm]{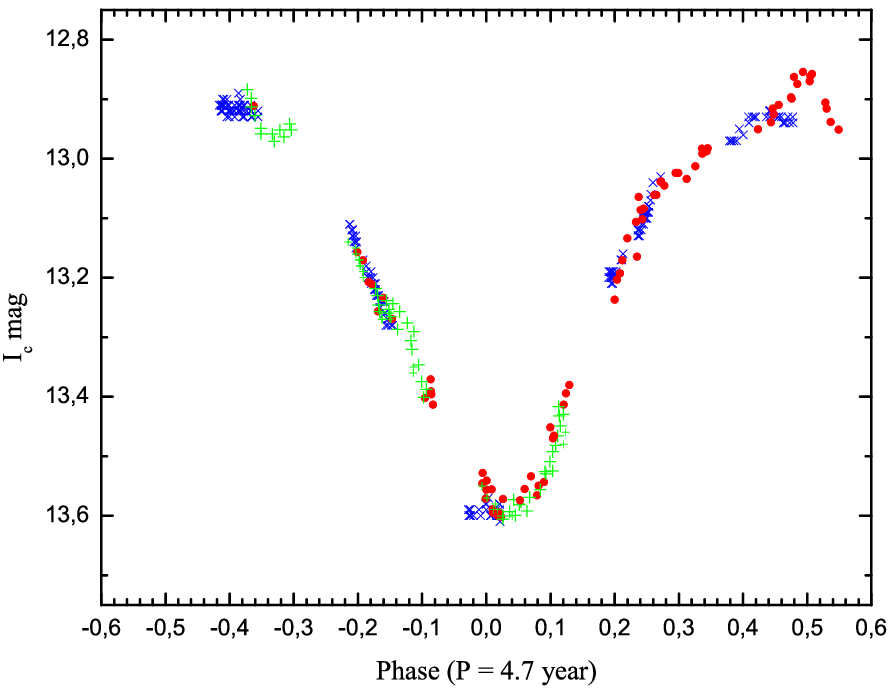}
\caption{{\it Top panel:} Light curves of \mbox{V718 Per} from the $I$-band to 5.8
$\mu$m including data from {\it Spitzer}, Grinin et al. (\cite{griB06}) and from
Cohen (\cite{coh04}) corrected to the Johnson-Cousins system. The arrows indicate
the times of spectroscopic observations. {\it Middle panel:} $I$-band
observations from Cohen (\cite{coh04}) (blue, $\times$), Nordhagen et al.
(\cite{nor06}) (green, $+$), and the present study (red, $\bullet$). {\it Bottom
panel:} The $I$-band data folded with a 4.7 year period.  {\it A full-color
version of this figure is available in the electronic version of this paper.}}
\label{fig:phot}
\end{figure}

\begin{figure}
\centering
\includegraphics[width=6cm]{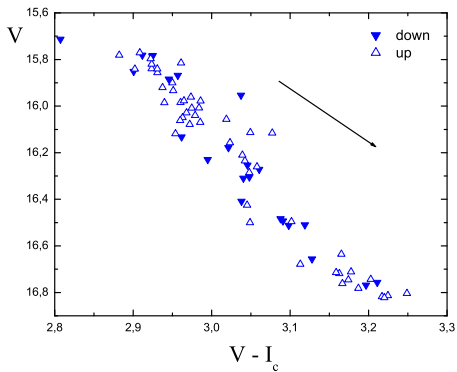}\\
\caption{The color-magnitude diagram $V$ versus $(V-I)$.
Filled triangles are points obtained when the star is fading in light
and open triangles when it is brightening.
The arrow indicates the standard reddening law. }
\label{fig:red}
\end{figure}

\section{Observations}

\subsection{Photometry}

We observed the entire 2005--2007 eclipse of \mbox{V718 Per} with $v,$ $r$ and $i$-band
photometry with the AZT-8 (0.7m) telescope of the Crimean Astrophysical
Observatory, Ukraine. After standard image reduction, the obtained aperture
photometry was transformed to the Johnson-Cousins $V,$ $R$ and $I$-bands. Since
\mbox{V718 Per} is located in a star forming region, almost all the stars in the field
exibit small brightness fluctuations. We selected three reference stars in the
field (H139, H205, and H 210) that show only small brightness variations ($\sim
0.05^{\rm m}$). They were calibrated via the photometric data of the stars TJ32,
TJ36 and TJ68 derived by Trullols and Jordi (\cite{tru97}). We estimate the
accuracy of the obtained photometric data to be about $0.03^{\rm m}$ in $V$ and
$R$, and $0.02^{\rm m}$ in $I$.  The photomeric data extend the series presented
earlier by Grinin et al. (\cite{griB06}).

Photometric observations of \mbox{V718 Per} in the near-IR wavebands $J$, $H$ and $K$
were obtained with the SWIRCAM camera of the 1.1-m telescope of the Pulkovo
Astronomical Observatory located at the Campo Imperatore Observatory, Italy.
These CCD images were reduced with standard reduction techniques, including bad
pixel removal, flat-field correction and sky subtraction. The differential
calibration of the photometric system with respect to the Johnson system was
performed with the help of the same reference stars as we used for the $I$-band.
The final accuracy of the $J$, $H$ and $K$-band data is about $0.02^{\rm m}$.

\mbox{V718 Per} was also observed in the 3.6, 4.5, 5.8 and 8.0\,$\mu$m bands
with IRAC on the Spitzer Space Telescope in February 2004, September 2004, and
October 2005 (3.6 and 5.8 $\mu$m only). Data from February 2004 were presented
by Lada et al. (\cite{lad06}) and Muench et al. (\cite{mue07}), who also
determined a limiting magnitude of $5.31$ in the 24\,$\mu$m MIPS band. We
retrieved all three datasets from the data archive of the Spitzer Science Center
and determined the fluxes of \mbox{V718 Per} in the four IRAC wavebands using
the aperture photometry tasks of the SExtractor package (Bertin \& Arnouts
\cite{ber96}). We used an aperture radius of 10 pixels and zero-point fluxes
recommended by the Spitzer Science Center (Reach et al. \cite{rea05}). For the
dataset from February 2004 we find for \mbox{V718 Per} 3.6 and 4.5 $\mu$m-band
magnitudes that are consistent with those presented by Lada et al.
(\cite{lad06}); the 5.8 and 8.0 $\mu$m-band magnitudes are slightly brighter,
most likely because of differences in the technique used to subtract the complex
background around \mbox{V718 Per}. We list our measurements in Table
\ref{tab:spitzer}.

\begin{table}
\caption{IR magnitudes derived from {\it Spitzer} observations of \mbox{V718
Per}}
\begin{tabular}{l c c c c}
\hline
\hline
Date        & $3.6\,\mu$m       & $4.5\,\mu$m       & $5.8\,\mu$m       & $8.0\,\mu$m\\
\hline
2004-02-11  & $9.37 \pm 0.02$   & $9.26 \pm 0.02$   & $9.01 \pm 0.02$   & $8.32 \pm 0.03$\\
2004-09-08  & $9.46 \pm 0.02$   & $9.39 \pm 0.02$   & $9.17 \pm 0.02$   & $8.56 \pm 0.02$\\
2005-10-06  & $9.62 \pm 0.02$   & --            & $9.37 \pm 0.04$   & --\\
\hline
\end{tabular}
\label{tab:spitzer}
\end{table}

\subsection{Spectroscopy}

Two high-resolution ($R = 45\,000$) spectra of \mbox{V718 Per}, covering $4800$ --
$8700$ {\AA}, were obtained with the HIRES spectrograph of the 10-m Keck
telescope on November 23, 2005 and December 10, 2006. The timing of the
observations is such that the first spectrum was obtained close to the eclipse
minimum (hereafter called the in-eclipse spectrum), while the second spectrum
was obtained about a quarter of a phase later (hereafter called the
out-of-eclipse spectrum). Although \mbox{V718 Per} was considerably brighter in 2006,
the out-of-eclipse spectrum has a lower signal-to-noise due to the poor weather
conditions at the time.

\section{Results}

An overview of the photometric observations of \mbox{V718 Per} from the $I$-band to 5.8
$\mu$m is presented in the upper panels of Fig.~\ref{fig:phot}. $I$-band data
collected so far, and the additional $V$ and $R$ data, show that the last
eclipse was very similar to the preceding one, which is demonstrated for the
$I$-band in the lower panel of  Fig.~\ref{fig:phot}, where the two eclipses are
folded with the 4.7 year period. Only minor deviations occur from the average at
a given phase suggesting that the eclipses are caused by the same
structure in the line-of-sight. Furthermore, the shape of the eclipses is almost
symmetric  around the centre of the eclipse, except for a small deviation at the
end of the last eclipse.

\mbox{V718 Per} becomes redder with decreasing brightness as demonstrated in the $V$
versus $(V-I)$ diagram shown in Fig. \ref{fig:red}, where also the reddening
line expected from obscuration by interstellar type grains is drawn. The colour
changes are similar during states of increasing and decreasing brightness.

\subsection{Spectral properties and photospheric parameters}

\begin{figure}
\centering
\includegraphics[angle=00, width=8.5cm]{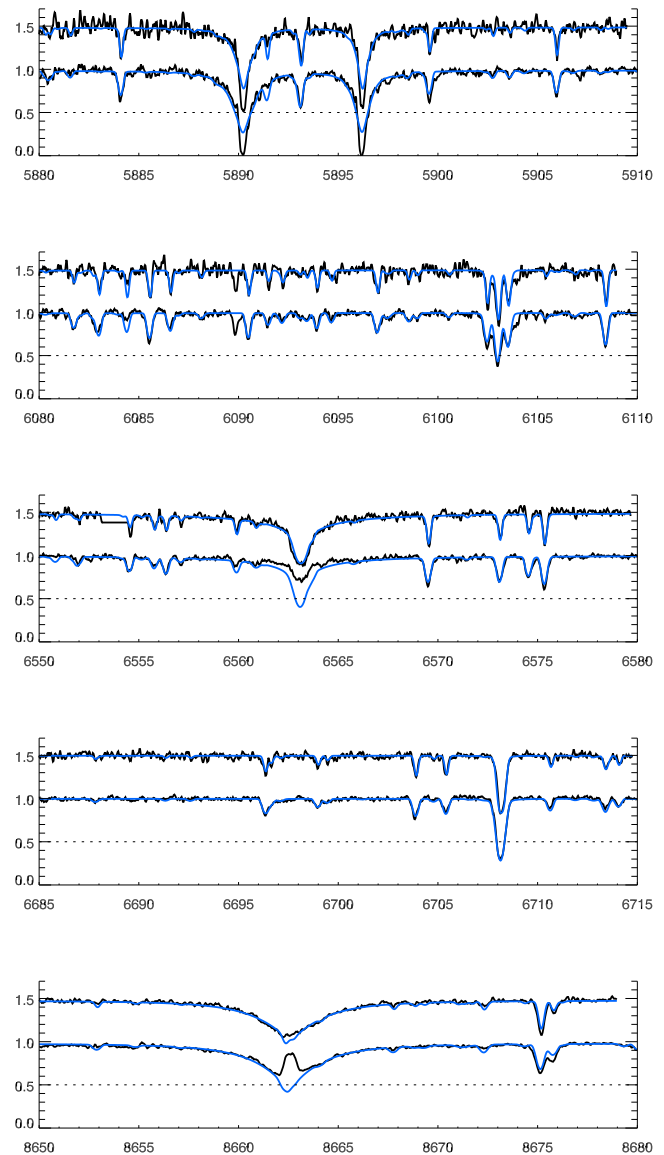}
\caption{Selected regions of the spectrum of \mbox{V718 Per}. In each panel we show
continuum-normalized profiles from the in-eclipse spectrum, as well as the
out-of-eclipse spectrum (offset with $+0.5$). Superimposed on each spectrum is a
synthetic model spectrum (blue). The 30 {\AA} wide regions cover from
top to bottom: the strong Na D absorption lines; a number of metallic lines
around 6100 {\AA}; H$\alpha$; the strong line of Li I at 6707 {\AA}; and one of
the Ca {\sc ii} IR triplet lines at 8663 {\AA}. {\it A full-color version of this
figure is available in the electronic version of this paper.}}
\label{fig:sp}
\end{figure}

\mbox{V718 Per} has a typical late-type spectrum without any strong emission lines. The
strong absorption line of Li~{\sc i} at 6708~{\AA} confirms its young age.
However, while the in- and out-of-eclipse spectra are rather similar, there are
a number of important differences. The most conspicuous is that the in-eclipse
spectrum shows weak core emission in the strong absorption lines of H$\alpha$
and the Ca {\sc ii} IR triplet. In addition, the absorption lines appear broader
in the in-eclipse spectrum. A large number of diffuse interstellar bands (DIBs)
are visible in the spectrum of \mbox{V718 Per}, most notably the DIBs at 6613 and 6379
{\AA}, but a number of weaker DIB features are present as well. These features
indicate large foreground extinction to \mbox{V718 Per}. There is no evidence that the
bands are strengthened in the in-eclipse spectrum, and we have no trace of any
other atomic or molecular features that could be associated with the occulting
object. Selected spectral regions of both spectra are plotted in Fig.
\ref{fig:sp}.

Although \mbox{V718 Per} is a faint star for high-resolution spectroscopy ($V =
15.8^{\rm m}$ in its brightest state), the spectra obtained with the 10-m Keck
are of sufficient quality to perform a spectroscopic analysis and determine the
fundamental photospheric parameters. We used the IDL-based package {\it
Spectroscopy Made Easy} (SME; Valenti \& Piskunov \cite{val96}) for the
calculation of synthetic spectra from model atmospheres and spectral line lists.
SME also includes a least-squares algorithm that can solve for the
stellar parameters that provide the best agreement with the observed spectrum,
most importantly the effective temperature ($T_{\rm eff}$), surface gravity
($\log g$), metallicity ($[M/H]$), projected rotational velocity ($v \sin i$)
and radial velocity ($v_{\rm rad}$).

In our analysis we used three spectral regions, together allowing us to
determine these essential parameters.
These regions include the following features: (1) Na {\sc i} D 5890 {\AA},
sensitive to both $T_{\rm eff}$ and $\log g$, (2) H$\alpha$ 6563 {\AA} and its
broad wings, mainly sensitive to $T_{\rm eff}$ and  (3) a large number of metal
lines around 6100 {\AA}, providing constraints on $[M/H]$, $v \sin i$ and
$v_{\rm rad}$. For our calculations we assumed a microturbulence $v_{\rm mic} =
1.0$ km s${}^{-1}$ and a macroturbulence of $v_{\rm mac} = 4.0$ km s${}^{-1}$,
both typical values for \mbox{G--K} type pre-main-sequence stars. Line data for
these calculations were taken from the VALD atomic line database (Piskunov et
al. \cite{pis95}; Kupka et al. \cite{kup99}), and model atmospheres from MARCS
(Gustafsson et al. \cite{gus03}).

We determined the fundamental photospheric parameters using all three wavelength
ranges simultaneously, but independently for the in- and out-of-eclipse spectra.
The results of these calculations are listed in Table \ref{tab:sme}.
One of the resulting synthetic spectra is plotted in Fig. \ref{fig:sp}
for comparison with the observed spectra.

Our analysis suggests a small difference in effective
temperatures, 5100\,K versus 5350\,K between in-eclipse and out-of-eclipse. This
difference is close to the $3\sigma$-limit, but the change is also
supported by the change in the Na {\sc i} D line profiles, as well as by
changes in temperature-sensitive absorption line depth ratios. Our values
of $T_{\rm eff}$ and $\log g$ derived for the out-of eclipse spectrum correspond
to spectral class G9V or IV according to the calibrations
by e.g. Kenyon \& Hartmann (\cite{ken95}) and Cohen \& Kuhi (\cite{coh79}).
The range in temperature is relatively small, and corresponds to one subclass
in terms of spectral type (G9--K0).

From the two spectra we measured an equivalent width of $385 \pm 15$ m{\AA}
(in-eclipse) and $315 \pm 15$ m{\AA} (out-of-eclipse) for the Li {\sc i} 6708
{\AA} line, which for a star with a temperature of $T_{\rm eff} \approx 5000$~K
and a gravity of $\log g \approx 3.5$ corresponds to a lithium abundance of
$\log n(\rm Li) \approx3.2$--$3.5$, indicative of a primordial abundance
(Pavlenko \& Magazz\`u \cite{pav96}).

An important difference between the two spectra is that the in-eclipse spectrum
shows stronger line broadening in the photospheric absorption lines than the
out-of-eclipse spectrum. This is not an instrumental effect, since the spectra
were obtained with an identical slit width. This is confirmed by the fact that
the narrow telluric absorption components have identical line widths in both
spectra.

\subsection{Radial velocities}

Accurate measurements of the radial velocity of \mbox{V718 Per} are important for
determining the true nature of this object, especially since one working
hypothesis is that \mbox{V718 Per}, in analogy with KH 15D, is a binary system. In
order to measure the stellar radial velocity as accurately as possible, we first
corrected instrumental drifts using telluric lines. The corrections recovered in
this way were small. We then shifted the spectra to the heliocentric reference
frame and determined from each spectrum the radial velocity of \mbox{V718 Per} by
cross-correlating the position of 250 well-isolated absorption lines with their
tabulated values.

The resulting heliocentric radial velocities are given in Table~\ref{tab:sme}. 
The average velocity of 14.5 km s$^{-1}$ is close to the average cluster
velocity of 14.0 km s$^{-1}$ (Kharchenko et al. \cite{kha05}). The difference
between the two measurements is small, only $74 \pm 80$ m s$^{-1}$. In other
words, we found no detectable change in radial velocity between the two spectra.
This is remarkable, since the first spectrum was obtained close to the deepest
point of the eclipse, and the time elapsed between the two spectra corresponds
to slightly less than a quarter of the 4.7-year period of eclipses. If \mbox{V718 Per}
were an edge-on binary system such timing would be particularly sensitive to changes in
radial velocity due to orbital motion. {\it Our non-detection of any change in
radial velocity therefore suggests that \mbox{V718 Per} is a single system.}

\begin{table}
\centering
\caption{Photospheric parameters derived for \mbox{V718 Per}}
\begin{tabular}{l r@{ $\pm$ }l r@{ $\pm$ }l}
\hline
\hline
Parameter & \multicolumn{2}{c}{In-eclipse} & \multicolumn{2}{c}{Out-of-eclipse}\\
\hline
$T_{\rm eff}$   & $5100$ & $100$ K      & $5350$ & $100$ K    \\
$\log g$    & $3.7$  & $0.1$    & $3.7$  & $0.1$      \\
$[M/H]$     & $0.1$  & $0.2$    & $0.1$  & $0.2$      \\
$v \sin i$  & $10.1$ & $0.2$ km s$^{-1}$ & $6.1$  & $0.2$ km s$^{-1}$ \\
$v_{\rm rad}$   & $14.504$ & $0.051$ km s$^{-1}$ & $14.578$ & $0.060$ km s$^{-1}$\\
\hline
\end{tabular}
\label{tab:sme}
\end{table}

\subsection{Stellar parameters}

The distance to IC 348 is usually assumed to be ${\sim}320$ pc, but has not been
determined very precisely due to variable extinction conditions over the cluster
(see the discussion by Herbig \cite{her98}). Even modern estimates based on {\it
Hipparcos} parallaxes differ; $260 \pm 25$ pc given by Scholz et al.
(\cite{sch99}) and $394$ pc by Kharchenko et al. (\cite{kha05}). In the following
we will adopt a distance of $300$ pc to \mbox{V718 Per}.

\begin{figure}
\centering
\includegraphics[angle=90, width=8.5cm]{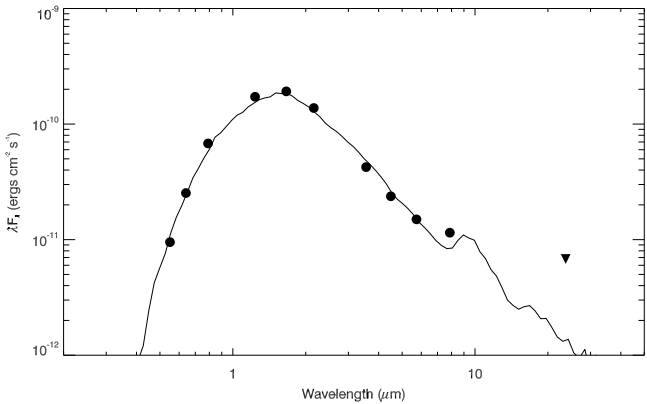}\\
\caption{Observed energy distribution of \mbox{V718 Per} as derived from photometry
data outside the eclipse phase (points) fitted to a model spectral energy
distribution (full-drawn curve), from which we derive basic parameters for the
star. The triangle indicates the upper limit to the flux at 24\,$\mu$m.}
\label{fig:sed}
\end{figure}

In order to derive the
extinction to \mbox{V718 Per} outside the eclipse phase and basic stellar
parameters we have made use of the program developed by Robitaille et al.
(\cite{rob07}), where we matched the observed out-of-eclipse energy distribution
to a set of model spectral energy distributions extracted from model atmosphere
calculations for young stars. The spectral energy distribution that shows the
best agreement with the observed fluxes and (spectroscopic) temperature is shown
in Fig. \ref{fig:sed}.
This best-fitting model has an effective surface temperature of
${\sim}5300$\,K and an extinction of $A_V = 4.7$.
The total
integrated stellar luminosity (assuming a distance of 300 pc) is then $3.4\,{\rm
L}_{\sun}$. Placing the star in the HR-diagram and comparing its position to
evolutionary model tracks by for instance Palla \& Stahler (\cite{pal99}) and
Siess et al. (\cite{sie00}) indicates that \mbox{V718 Per} is in its radiative
contraction phase with a mass of ${\sim}1.6\,{\rm M}_{\sun}$
and an age of ${\sim}5$ Myr. The IC 348 cluster members
were found to show an age spread of between $0.5$ and $10$ Myr by Luhman et al.
(\cite{luh98}), but the pre-main sequence stars peak at 2.5 Myr with an age
spread of 4 Myr according to Muench et al. (\cite{mue07}). From its location in
the HR diagram, and also its spectral properties, we conclude that \mbox{V718
Per} is a post-T Tauri-star and among the older pre-main-sequence objects in the
region. This is also supported by the lack of significant line emission, and the
weak IR excess emission which indicates that \mbox{V718 Per} has only a thin,
low-mass disk.

\section{Discussion}

Our observations shed some new light on the possible nature of \mbox{V718 Per}. Its
location above the main sequence in the HR-diagram and the large Li abundance
derived confirm the pre-main-sequence nature of the object. We have found that
\mbox{V718 Per} is in its post-T Tauri phase, and in accordance the spectrum lacks
emission lines, for instance of H$\alpha$. Since the last two eclipses have very
similar light curves, the eclipses are likely to be caused by the same
obscuring structure. Presumably this extended feature is part of a disk observed
at high inclination, possibly edge-on. The disk produces the observed IR excess,
but the absence of an IR excess in the $JHK$ bands suggests that there is a gap
with a radius of a few AU inside the disk edge. The extended circumstellar
feature could represent an irregular azimuthal mass distribution in the disk, or
a warped disk structure. The feature could also be related to spiral arm
structures that developed because of perturbations from a secondary
companion, as discussed by Sotnikova \& Grinin (\cite{sot07}).

It has been speculated that \mbox{V718 Per} is a binary, but from our
high-resolution spectra we find no detectable change in radial velocity.
If \mbox{V718 Per} is indeed experiencing occultations from its circumstellar
disk, it is probably correct to assume that the systems is observed edge-on and
it is therefore most likely a single star and not similar to KH 15D. From
the eclipse period of 4.7 years and a stellar mass of $1.6~{\rm M}_{\sun}$ (see
Section 3.4), the occulting structure is orbiting at a distance of about 3.3 AU
from the star, presumably at the inner disk radius. The presence of a central
gap can be a natural consequence of \mbox{V718 Per} being in a relatively
evolved pre-main-sequence phase of evolution, but a stable irregularity in the
disk may flag the presence of a perturbing low-mass object, for instance in 1:1
resonance motion
(see Ozernoy et al. \cite{oze00}). From the limits on any radial velocity
change of the star (Table \ref{tab:sme}), we find that the mass of any planet
orbiting close to the disk plane and inside the disk edge cannot exceed
$6~M_{\rm Jup}$. A warped disk edge could in principle be maintained by a planet
in an inclined orbit,  in which case the perturbing object can be more massive.

\begin{figure}
\centering
\includegraphics[angle=90, width=8.5cm]{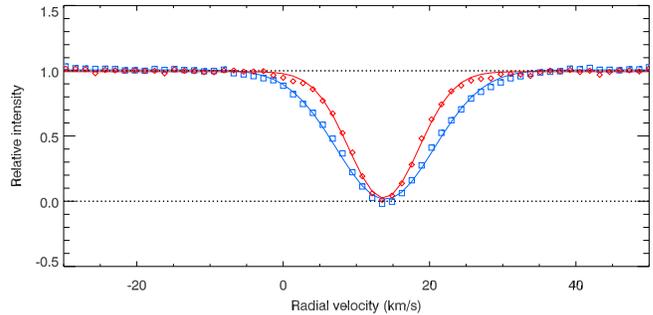}\\
\caption{ The ``best-average'' absorption line profiles reconstructed with the
technique of least-squares deconvolution (LSD - see text), plotted in the
heliocentric reference frame. The (narrow) in-eclipse profile is shown as red
diamonds, and the (wide) out-of-eclipse profile as blue squares. For each set of
points a Gaussian curve that best fits the data is overplotted. No rescaling has
been applied to the individual profiles. {\it A full-color version of this
figure is available in the electronic version of this paper.}} 
\label{fig:profile}
\end{figure}

When \mbox{V718 Per} declines in brightness, it becomes redder, and we found evidence
that the star becomes somewhat cooler. At the same time the absorption lines
become broader. We investigated the exact shape of the absporption line profiles
using the technique of least-squares deconvolution (LSD, see Donati et al.
\cite{don97}), which allows us to study even the smallest changes in the shape
and depth of spectral lines. LSD reconstructs a ``best-average'' line profile by
numerically combining the shape of a large number of absorption lines. For our
spectra of \mbox{V718 Per}, we reconstructed these ``best-average'' absorption line
profiles using the same narrow and isolated absorption lines as we used for the
determination of radial velcocities. The reconstructed profiles clearly show
that the in-eclipse profiles are substantially broader than the out-of-eclipse
profiles, and that the out-of-eclipse profiles are only slightly shallower than
the in-eclipse profiles (Fig. \ref{fig:profile}). Hence, the photospheric
absorption lines increase in equivalent width with decreasing brightness in
accordance with the change in $T_{\rm eff}$. Moreover, some lines show emission
reversals at minimum light (see Fig. \ref{fig:sp}). The spectrum of \mbox{V718 Per} is
therefore rather different in the two phases observed.

The Ca {\sc ii} and H$\alpha$ emission components seen in the in-eclipse
spectrum are too strong to be explained entirely as an effect of changing the
continuum-to-line contrast between the in- and out-of-eclipse spectra. Such an
effect would arise if for instance the emission region is extended and becomes
less occulted than the stellar disk by a sharp dust screen in the foreground.
The emission components are central and relatively narrow and are likely to
originate close to the star. We conclude that the stellar photosphere and an
associated emission region has changed between the time the two spectra were
taken. With only two spectra we cannot speculate on how this change comes about.
It could be related to a larger surface coverage of dark stellar spots, which
happened to occur during the time the in-eclipse spectrum was exposed. This
would explain the observed small shift towards cooler surface temperatures and
also the enhanced ``chromospheric'' emission. There is no trace of any
rotational modulation from spots in the light curves. A variable degree of spot
activity must then be confined to the polar regions of the star.

However, stellar surface spots cannot explain the observed increase in the
broadening of the photospheric absorption lines. The changes in the
absorption lines also cannot be explained by line-doubling from a binary
system, since that would conserve the total equivalent width. Instead, we
propose that this line broadening is a consequence of increasing light
scattering when the thick part of the disk passes the line-of-sight to the star.
The Keplerian velocity of a circumstellar feature with a $4.7$-year
period is ${\sim}21$ km s$^{-1}$, which is sufficient to broaden the absorption
line profiles by \mbox{${\sim}4$ km s$^{-1}$.} Since the circumstellar feature must
extend over at least half a circle along the disk edge, scattering must be
anisotropic, with a strong preference of forward scattering. Similar line
profile changes resulting from scattering in T Tauri disks have been modelled by
Grinin et al. (\cite{griM06}) assuming similar asymmetric scattering functions.

More observations of \mbox{V718 Per} are warranted. Continued photometric and
spectroscopic monitoring can establish more firmly the stability of the extended
dust feature and also how the photospheric and chromospheric spectral features
vary with time. V718~Per resembles the UX Ori stars, earlier type stars that
undergo usually more irregular occultations by dusty blobs (see Grinin
\cite{gri00}). In analogy with these objects, we also expect the light from V718
Per to be polarized, and that the degree of linear polarization should increase
drastically with decreasing brightness, which can be tested by observations.
Finally, more information on the mid-IR and sub-mm fluxes is needed in order to
probe and constrain the distribution and mass of the circumstellar material
around \mbox{V718 Per}.

\section{Conclusions}

Our new photometric data of \mbox{V718 Per} (H187) extends previous measurements and
confirms that this object shows long-lasting eclipses with a period of 4.7
years. The eclipses, which are very symmetric,  are caused by occulting dust,
and the colour changes suggest an extinction law of the foreground dust that
deviates only little from what is expected for ``normal'' interstellar grains.

It has been speculated that \mbox{V718 Per} may be similar to the unusual close
binary KH 15D, also showing periodic eclipses, which presumably are caused by
different coverage of the orbiting stars by a circumbinary disk. We
obtained two
high-dispersion spectra of \mbox{V718 Per}, the first
close to the deepest
point of the eclipse and the other at a time outside eclipse with a time
difference corresponding to roughly a quarter of the eclipse period. 
From these spectra we found no evidence of any change in the radial
velocity of the star to within $\pm 80$ m s$^{-1}$. Although we cannot be
fully certain on the basis of two spectra, the absence of radial velocity
variations in a system that experiences eclipses from its circumstellar material
makes it very unlikely that \mbox{V718 Per} is a close binary. \mbox{V718 Per}
seems therefore very different from KH 15D.

We confirm that the eclipses are caused by a stable extended, dusty structure
orbiting at ${\sim} 3.3$ AU from the star in an edge-on circumstellar disk. We
have derived the SED of \mbox{V718 Per} outside eclipse, including IR
fluxes obtained with {\it Spitzer}. This indicates that the star could be
surrounded by a thin, low-mass disk.
Because of the extended eclipse duration, the structure that causes the
eclipses must extend over half a circle along the disk edge. A
low-mass companion can in principle induce this structure,
but the mass of any planet
cannot exceed $6~M_{\rm Jup}$ (assuming an edge-on system). The full
amplitude of an eclipse amounts to 1.1 magnitudes in the V-band, but there are
no signs of any enhanced absorption features from circumstellar gas during
eclipse. It appears that the occulting structure is rather void of gas. 

\mbox{V718 Per} has a typical late-type absorption line spectrum without strong
emission lines of e.g H$\alpha$. We have used theoretical synthetic spectra from
model atmosphere and derived fundamental photospheric parameters. Our
spectroscopic analysis shows
that \mbox{V718 Per} has a primordial abundance of Li and a surface temperature of
$T_{\rm eff} \approx 5200$\,K. From the luminosity derived from its SED, and by
comparing with theoretical evolution tracks, we find that \mbox{V718 Per} is in its
post-T Tauri phase of evolution.

However, there are remarkable differences between the in-eclipse and
out-of-eclipse spectrum. During the eclipse several spectral features show that
the surface temperature is slightly lower than in the out-of-eclipse spectrum,
corresponding to a change in spectral type from G9 to K0. In addition, narrow
emission components appear in the absorption cores of the H$\alpha$ and the Ca
{\sc ii} IR triplet lines during eclipse, and the photospheric absorption lines
become slightly broader. The change in stellar surface temperature and the emission
line activity observed is puzzling. {\it Since \mbox{V718 Per} shows no
short-term (rotational) photometric variability, this cannot be explained as the
result of a variable coverage by starspots.} However, the observed spectral
changes could be related to long-term changes in the activity near the polar
regions.
The broadening of the absorption lines we ascribe to forward scattering of
stellar light in the circumstellar dust feature when it passes through the
line-of-sight. 

\begin{acknowledgements}

We are greatly indebted to George Herbig for obtaining the two spectra of V718
Per with the HIRES spectrograph at the Keck Observatory. We also thank Valery
Larionov for his help in obtaining the optical photometry of \mbox{V718 Per} and Tanja
Ryabchikova and Yurij Pakhomov for their help with the spectroscopic analysis.
This work was supported in part by the program of the Russian Acad.
Sci.\,``Formation and Evolution of Stars and Galaxies'', grant N.Sh. 6110.2008.2
and the grant INTAS 03-51-6311. The W.M. Keck Observatory is operated as a
scientific partnership among  the California Institute of Technology, the
University of California, and the National Aeronautics and Space
Administration. 

\end{acknowledgements}

\end{document}